\documentclass[11pt,twoside]{article}

\usepackage{asp2006}
\usepackage{epsf}
\usepackage{lscape}
\usepackage{graphicx}
\begin{document}
\title{The near-IR shape of the big blue bump emission from quasars:
under the hot dust emission} 

\author{Makoto Kishimoto$^{1,2}$, Robert Antonucci$^3$, Omer Blaes$^3$}

\affil{$^1$Max-Planck-Institut f\"ur Radioastronomie, 53121 Bonn,
Germany\\ $^2$SUPA (Scottish Universities Physics Alliance), Institute
for Astronomy, University of Edinburgh, EH9 3HJ, UK\\ $^3$Physics
Department, University of California, Santa Barbara, CA 93106, USA}

\begin{abstract} 

One primary difficulty in understanding the nature of the putative
accretion disk in the central engine of AGNs is that some of its key
intrinsic spectral signatures are buried under the emissions from the
surrounding regions, i.e. the broad line region (BLR) and the
obscuring torus. We argue here that these signatures can be revealed
by using optical and near-IR polarization.  At least in some quasars,
the polarization is seen only in the continuum and is not shared by
emission lines. In this case, the polarized flux is considered to
show the intrinsic spectrum interior to the BLR, removing off the
emissions from the BLR and torus. We have used this polarization to
reveal the Balmer-edge feature and near-IR spectral shape of the
central engine, both of which are important for testing the
fundamental aspects of the models.

\end{abstract}


\section{The key, but hidden spectral signatures from the central engine}  

The radiative output of AGNs is dominated by the UV/optical component,
often called the big blue bump (BBB). This is usually attributed to be
from an accretion disk around a supermassive black hole, but the
nature of this putative disk has not been well understood in many
respects, i.e. there are unsatisfactory agreements between theory and
observations (e.g. Antonucci 1988; Koratkar \& Blaes 1999). However,
there is certainly a big difficulty in the observational side.  We
still do not have the spatial resolution to isolate the central engine
from the surrounding regions, most notably the broad-line region (BLR)
and the obscuring torus. Consequently, the central engine's spectrum
is always mixed with the emissions from the BLR and torus, and some of
the key, intrinsic spectral features of the central engine are almost
impossible to uncover.

The hot dust thermal emission from the torus starts to dominate
longward of $\sim 1 \mu$m (set by a dust sublimation temperature) and
essentially hide the near-IR part of the BBB. This wavelength region
is actually quite important, since disk models make a few key,
well-known predictions at these long wavelengths.  The spectral shape
of a simple multi-temperature blackbody disk without any outer
truncation asymptotically becomes as blue as $f_{\nu} \propto
\nu^{+1/3}$ at long wavelengths. This is also true in more
sophisticated, bare-disk atmosphere models (e.g. Hubeny et al. 2000),
and the limit is essentially reached longward of 1 $\mu$m for
reasonable parameters of quasars.  Also, outer parts of standard disks
are known to be unstable against self-gravity, possibly leading to a
truncation at an outer radius, but this is predicted to be around a
near-IR emitting radius for quasar parameters (e.g. Goodman
2003). This would form a spectral break in the near-IR, making the
shape even bluer.  Thus, the near-IR part of the BBB is quite
important for robust tests of disk models.

Another key spectral region is the Balmer edge.  Among the opacity
edge features generally predicted by disk atmosphere models, the
Balmer edge has the advantage (over the Lyman edge) of being much less
prone to foreground absorptions.  However, high-order Balmer emission
lines and Balmer continuum (and also FeII blends) from the BLR heavily
contaminate the wavelength region, so that the Balmer edge feature
intrinsic to the BBB becomes essentially unobservable.

We argue here that we can get rid of these contaminating emissions
from the BLR and torus and extract the central-engine-only spectrum by
using optical and near-IR polarization.

\begin{figure}
\begin{center}
\includegraphics[width=9cm]{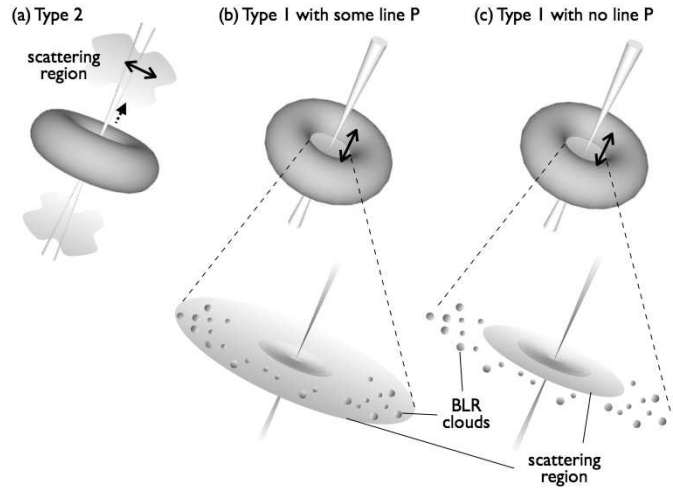}

\caption{Schematic diagrams for the geometry of dominant scattering
regions for (a) Type 2s (b) Type 1s with some line polarization (c)
Type 1s with no line polarization. In each panel, the double arrow
shows the position angle of continuum polarization.}

\end{center}
\end{figure}

\section{The polarization of Type 1 AGNs} 

The optical polarization which is probably the most familiar to many
people in the field is the one seen in Type 2 AGNs. The broad emission
lines are seen in the polarized flux in many of these, with
polarization position angle (PA) perpendicular to the radio jet axis
(Antonucci 1993). The interpretation is that the gas which resides
{\it outside} the BLR, along the jet axis perhaps above and below the
torus, scatters the light from the central engine {\it and} BLR into
the line of sight (Fig.1a). Thus they both show up in the polarized
flux.

What we want to utilize here is not the polarization in these Type 2s,
but that in Type 1s. In those objects, a different, nuclear
polarization component appears to dominate, quite plausibly and simply
because the bright nuclear region inside the torus is directly seen.
In many cases, the continuum is polarized typically at $P\sim$1\%
level, with the PA {\it parallel} to the jet axis (e.g. Berriman et
al. 1990).  In many Seyfert 1s, the broad lines are also polarized in
addition to the continuum, often at lower $P$ and at different PA than
continuum (it rotates across the line wavelengths; e.g. Smith et
al. 2004; see also Lira et al. in these proceedings). These imply that
the scattering region is more or less similar in size to the BLR. The
parallel polarization quite possibly indicates that the scattering
region is in a flattened/equatorial optically-thin geometry having its
symmetry axis along the jet direction (Fig.1b).

At least in some quasars, however, similar continuum polarization is
seen without any line polarization --- this is what we want to use.
Since lines aren't polarized, scattering is considered to be caused
{\it interior} to the BLR (Fig.1c), by electrons (since the site is
inside the dust sublimation radius). Then the polarized flux would in
fact be an electron-scattered copy of the intrinsic spectrum of the
central engine, with all the emissions from the BLR and torus
eliminated!

\begin{figure}[t]
\begin{center}
\includegraphics[width=11cm]{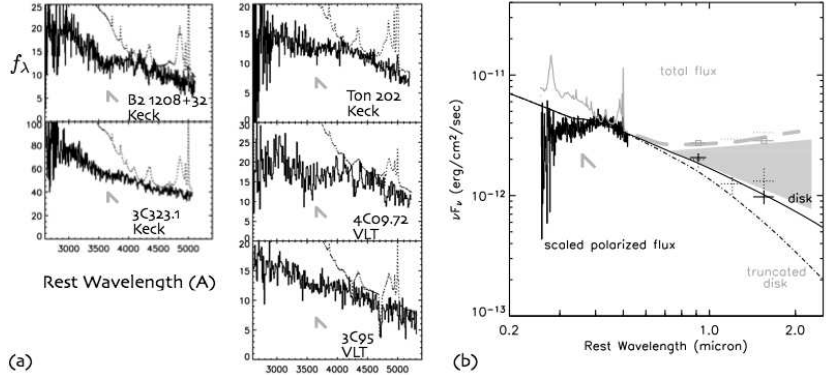}

\caption{(a) Optical spectropolarimetry of 5 quasars from Kishimoto et
al. (2004). The solid line is the polarized flux in units of
10$^{-18}$ ergs/cm$^2$/sec/\AA. The dotted line is the total flux
scaled to match the polarized flux at the red side. The wavelength of
the Balmer discontinuity, 3646\AA, is indicated as a folded line in
each panel. (b) The optical to near-IR SED for the quasar Ton202 in
$\nu F_{\nu}$ with both axes in log scale.  The top gray line is the
optical total flux which is schematically connected by a thick gray
dashed curve to the near-IR total flux shown in thin gray lines with
square symbols. The solid black line and crosses are the optical and
near-IR polarized flux, respectively, scaled to match the total flux
at the red side in the optical. The near-IR total and polarized flux
taken at a different epoch are shown in dotted lines and crosses.  The
gray wedge schematically shows the emission from the torus (plus
possibly the host galaxy) which is removed by measuring the polarized
flux.  A disk atmosphere model without any disk truncation is shown
in a smooth solid curve, while the dot-dashed curve is the same disk
which is truncated at the self-gravity-unstable radius.  From
Kishimoto et al. (2005).}

\end{center}
\end{figure}

\section{Revealing the Balmer edge and near-IR spectral shape}

We have first applied this idea for looking closely at the Balmer edge
wavelength region in several quasars with spectropolarimetry.  At the
H$\beta$ wavelengths, we essentially do not see any line features in
the polarized flux, which is the basis for the application of our idea
(Fig.2a). Then we find that the Balmer edge is seen in absorption:
there is a downturn at around 4000\AA\ and an upturn at around
3600\AA\ in the polarized flux spectra of all the five objects
(Fig.2a). We interpret this Balmer-edge feature to be intrinsic to the
central engine, indicating quite fundamentally that the big blue bump
is indeed from a thermal and optically-thick emitter.

Then we have also been trying to accurately measure near-IR polarized
flux to obtain dust-eliminated near-IR shape of the BBB.  Fig.2b shows
our first object, where the polarized flux is compared with the total
flux shape from the optical to near-IR (note that it is one of the
objects in Fig.2a).  The near-IR polarized flux appears to be nicely
removing the torus emission (and also perhaps the host galaxy light as
well).  The spectral index measured between $J$ and $K$ ($\lambda_{\rm
rest} = 0.9 - 1.6 \mu$m) in the polarized flux is $\alpha = +0.42 \pm
0.29$ ($f_{\nu} \propto \nu^{\alpha}$), which is intriguingly
consistent with the $\nu^{+1/3}$ limit.  A model atmosphere spectrum
without a disk truncation runs through optical to near-IR data points,
though it fails in the near-UV side (Fig.2b).  The data do not appear
to show evidence for a disk truncation, based on the model spectrum
for a disk truncated at the self-gravity-unstable radius.  Of course
we need more measurements to see if similar spectra are seen
systematically in other objects, which will check the validity of our
interpretation for the near-IR polarized flux.

\section{Conclusions}

It seems that these polarization measurements are starting to
delineate the fundamental aspects of the radiative output from the
central engine. The results so far suggest that the big blue bump
emitter at least resembles a standard disk in a few respects.  We hope
to explore its nature with further measurements.




\end{document}